# Programmable coherent linear quantum operations with high-dimensional optical spatial modes

*Shikang Li[1], Shan Zhang[1] Xue Feng[1,\*], Stephen M. Barnett[2], Wei Zhang[1], Kaiyu Cui[1], Fang Liu[1], Yidong Huang[1]*

\*Corresponding Author: E-mail: x-feng@tsinghua.edu.cn
[1]Department of Electronic Engineering, Tsinghua University, Beijing 100084, China.
[2]School of Physics and Astronomy, University of Glasgow, Glasgow G12 8QQ, United Kingdom.

*Abstract.*

A simple and flexible scheme for high-dimensional linear quantum operations on optical transverse spatial modes is demonstrated. The quantum Fourier transformation (QFT) and quantum state tomography (QST) via symmetric informationally complete positive operator-valued measures (SIC POVMs) are implemented with dimensionality of 15. The matrix fidelity of QFT is 0.85, while the statistical fidelity of SIC POVMs and fidelity of QST are ~0.97 and up to 0.853, respectively. We believe that our device has the potential for further exploration of high-dimensional spatial entanglement provided by spontaneous parametric down conversion in nonlinear crystals.

*Introduction*.

Photonics provides an outstanding platform for exploring non-classical computational resources [1] since the entanglement can be conveniently generated through optical nonlinear effects [2-4], while linear manipulation protocols are available in multiple degrees of freedom [5-7]. Great efforts have been made on high-dimensional entangled states, both for tests of quantum mechanics and also for applications to quantum technology [8]. There is a push to increase the information encoded on a single photon [9] and achieve high-dimensional universal linear operation to extend the capacity of quantum processing as well as enhance the versatility of quantum computing and simulation [10]. High-dimensional quantum encoding has been demonstrated on photons exploiting the domains of optical path, frequency and transverse spatial modes. For the first of these Reck *et.al*, showed how arbitrary unitary operators could be realized by cascaded basic blocks consisting of phase modulators and couplers [5]. In the Reck scheme, programmable $6 \times 6$ matrix operators [11] and projectors with dimension of 15 [8] have been reported, but high-dimensional path entanglement generation remains a challenge [8]. In frequency domain, the entanglement can be generated routinely through spontaneous four wave mixing (SFWM) in optical fibers or silicon waveguides [4], but the achieved dimensionality of quantum operation is limited merely to $3 \times 3$ as ultra-fast electric optic modulation (EOM) devices are required [7]. To exhibit substantial computational superiority, the dimensionality has to be increased.

Transverse spatial modes provide abundant resources for quantum encoding and processing. Spatial mode entanglement induced by spontaneous parametric down conversion (SPDC) in nonlinear crystals has been well investigated [2,3,12,13]. They have been employed for ghost imaging [14] and entangled qudits generation [15,16]. Though spatially entangled photon pairs with ultra-high Schmidt numbers [15,17] can be readily obtained through SPDC, the exploitation of such quantum resources has been hindered due to the lack of universal operation protocols. Quantum operation protocols on orbital angular momentum (OAM), which can be considered as a specific basis of spatial modes, has been presented but only with limited dimensionality[6]. Thus, although high-dimensional universal quantum operations on spatial modes could boost the exploration of currently existing entanglement resources, an efficient approach remains to be demonstrated.

Here, we tackle this issue by proposing and demonstrating a distinctive method to coherently manipulate spatial modes that are arbitrarily distributed in the transverse plane according to the beam propagation. Quantum operators have been experimentally achieved with dimensionality up to $15 \times 15$, and we apply these to demonstrate a quantum Fourier transform (QFT) and symmetric informationally complete positive operator-valued measures (SIC POVMs). This is, to the best of our knowledge, the implementation of discrete universal quantum operators with the highest reported dimensionality [6-8,11]. Due to the universality, precision, and controllability, our scheme would be possible to fully explore the whole spontaneous down-conversion cone [2] for high-dimensional demonstrations of nonclassical phenomena.

*Principle.*

Generally, with a set of well-defined orthogonal basis, any linear operation can be expressed as $|\beta\rangle = T|\alpha\rangle$, in which $T$ is a complex matrix and $|\alpha\rangle$ and $|\beta\rangle$ are initial and final state vectors in a complex vector space with dimensionality of *N*. In our previous works [18,19], the state vector of $|\alpha\rangle$ and $|\beta\rangle$ can be encoded on a set of orthogonal discrete spatial modes $|\varphi_n\rangle = u(\bm{r}-\bm{R}_n)$. As indicated in Fig.1 (a), $\bm{R}_n$ is the transverse coordinate of the *n*-th discrete spatial mode while $u(\bm{r}-\bm{R}_n)$ is considered as Gaussian function of $u(\bm{r}-\bm{R}_n) \propto \exp(-|\bm{r}-\bm{R}_n|^2/w_0^2)$ and the optical waist of $w_0$ is designed as sufficiently small compared with transverse distance $|\bm{R}_n - \bm{R}_m|$ between two spatial modes to ensure the orthogonality [19]. Hence the state vector can be expressed as $|\alpha\rangle = \sum a_n |\varphi_n\rangle$. Specifically, such discrete spatial modes encoded qudits can be treated as the generalized expression of quasi-OAM-encoded qudits and path-encoded qudits, since the distribution of transverse coordinates $\{\bm{R}_n\}$ can be chosen arbitrarily and designed at will. For example, with 3D waveguide technology [], integrated path-encoded photonic qudits can be coupled to our linear transformation scheme with considerable flexibility and efficiency.

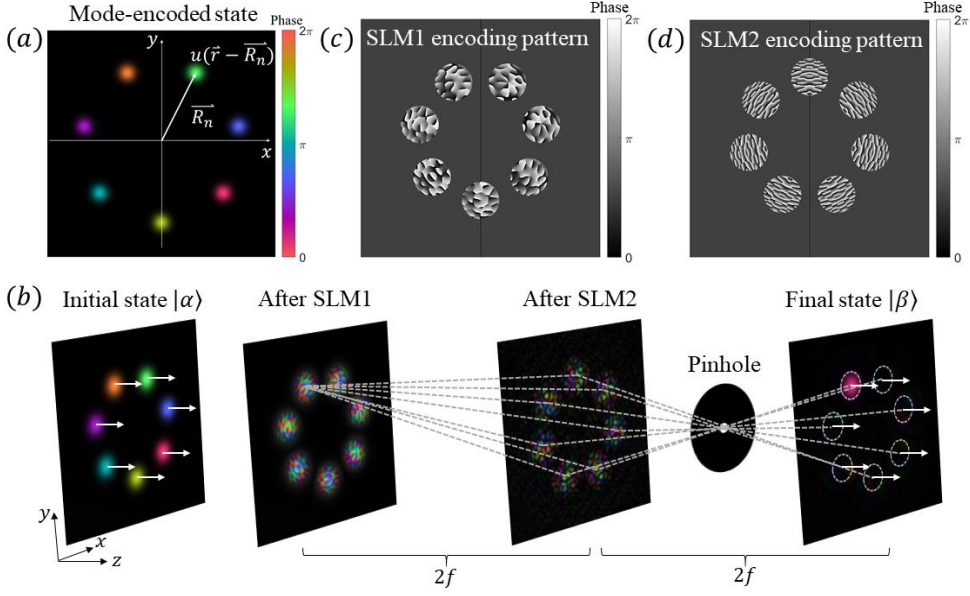

Fig.1 (a) Definition of the discrete spatial modes. (b) Schematic setup for implementing linear quantum operators. A 7-dimensional QFT demonstration is shown as a concrete example. Simulated field evolutions after each SLM are presented. Phase modulation pattern implemented on SLM1(c) and SLM2 (d) for beam splitting and recombining, respectively.

For a given linear operator of $T$, one-to-$N$ beam splitting and $N$-to-one beam recombining are required to achieve the columns and rows of a general transformation matrix, *i.e.*, $T$ should be decomposed as the Hadamard product of two matrices ($T_{mn} = A_{mn} B_{mn}$). The optimal decomposition to achieve maximum energy efficiency has been discussed in our previous work [19]. With the obtained matrices of A and B, beam splitting and recombining can be implemented by two phase-only spatial light modulators (SLM1 and SLM2) combined with two 2$f$ systems and a pinhole (Fig.1 (b)). Thus, the key question is how to determine the phase modulation pattern on each SLM. Considering a diffraction grating illuminated by an optical beam, the diffraction on the Fourier plane is the convolution of Fourier coefficients of the grating and Fourier spectrum of the incident beam field according to diffraction theory [20]. Thus the diffraction gratings on SLM1 and SLM2 should be set according to the Fourier coefficients of $A_{mn}$ and $B_{mn}$:

$$G_{1n}(\boldsymbol{r}) = \sum_{m=1}^{N} A_{mn} \exp\left[ i\boldsymbol{k}_{mn} \cdot (\boldsymbol{r} - \boldsymbol{R_n}) \right]$$
$$G_{2m}(\boldsymbol{r}) = \sum_{n=1}^{N} B_{mn} \exp\left[ -i\boldsymbol{k}_{mn} \cdot (\boldsymbol{r} + \boldsymbol{R_m}) \right],$$

(1)

where $\boldsymbol{k}_{mn}$ is transverse wave vector expressed as $\boldsymbol{k}_{mn} = k(\boldsymbol{R_n} - \boldsymbol{R_m})/2f$. Eq. (1) denotes ideal modulation for beam splitting and recombining applied to $|\varphi_n\rangle$ and $|\varphi_m\rangle$, respectively. However, to generate the diffraction gratings in Eq. (1), amplitude gain would be required, which is unachievable by passive modulations. Thus, to avoid the amplitude gain, the actual phase modulation functions for the beam splitting beam recombining are settled as:

$$H_{1n}(\boldsymbol{r}) = \exp\left\{i \arg\left(\sum_{m=1}^{N} \mu_{mn} A_{mn} \exp\left[i\boldsymbol{k}_{mn} \cdot (\boldsymbol{r} - \boldsymbol{R}_n)\right]\right)\right\}$$
$$H_{2m}(\boldsymbol{r}) = \exp\left\{i \arg\left(\sum_{n=1}^{N} v_{mn} B_{mn} \exp\left[-i\boldsymbol{k}_{mn} \cdot (\boldsymbol{r} + \boldsymbol{R}_m)\right]\right)\right\} \quad (2)$$

The amplitudes of the diffraction gratings are restricted to 1. As a concrete example, the diffraction gratings for $7 \times 7$ QFT are displayed in Fig. 1 (c) and (d). Two sets of undetermined coefficients of $\{\mu_{mn}\}$ and $\{v_{mn}\}$ are introduced to achieve desired beam splitting and recombining ratio. To determine $\{\mu_{mn}\}$ and $\{v_{mn}\}$, the gradient descent algorithm is employed to maximize the fidelity between target and implemented matrix ($A$ or $B$ vs. $A_{\exp}$ or $B_{\exp}$) represented by Fourier coefficients of Eq. (2). The fidelity is normalized by energy as [21]:

$$Fide(A_{\exp}, A) = \left|\frac{Tr(A^\dagger A_{\exp})}{\sqrt{Tr(A_{\exp}^\dagger A_{\exp}) \cdot Tr(A^\dagger A)}}\right|^2, \quad (3)$$

Here only optimizing the phase grating of beam splitting is discussed, as that for the beam recombining is similar. An amplitude-flatted version of Eq. (1) can be obtained by directly setting $\{\mu_{mn}\} \equiv 1$. In Fig.2 (a) and (b), Huygens-Fresnel simulation results for beam splitting gratings of dimensionality up to 51 are summarized. Each data point is the statistical combination of 20 random complex target matrices. The data labeled as "Original" correspond to $\{\mu_{mn}\} \equiv 1$ and there is a slight fidelity deterioration as dimensionality grows, while unitary fidelity values could be achieved with small efficiency penalty (<2%) after optimization. The efficiency is calculated by $Effi(A_{\exp}, A) = Tr(A_{\exp}^\dagger A_{\exp})/Tr(A^\dagger A)$. The enlarged fidelity values after optimization are shown in the inset of Fig.2 (a). We believe that the imperfection is caused by the calculation error such as iteration tolerance. Similar results have been obtained on beam recombining. As shown in Fig. 2(c), the transverse coordinates of our proposed discrete spatial modes are arranged on a circle to mimic the spatially sampled Type-I SPDC cone. Compared with the recent breakthrough technology of multi-plane light conversion (MPLC) [], the achieved dimensionality of arbitrary transformations in this work is much higher. Besides, the cascaded structures are avoided, while the number of cascaded spatial modulation elements would grow significantly as the dimensionality grows with MPLC approach.

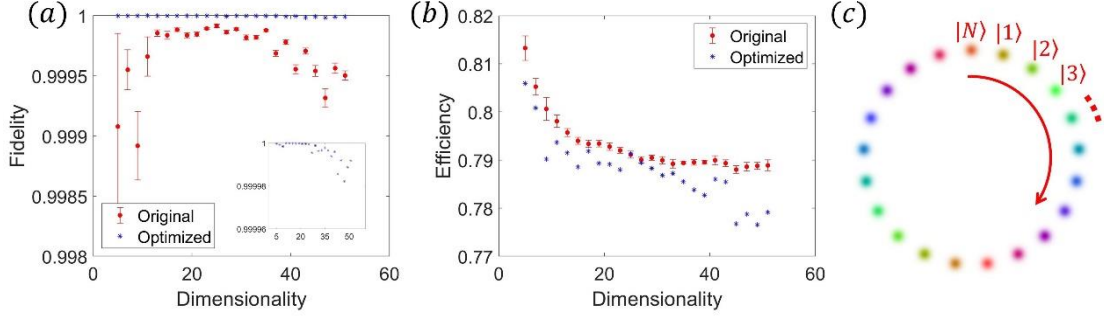

Fig.2 Simulated fidelity (a) and efficiency (b) for one-to-N beam splitting. (c) The transverse coordinates of employed spatial modes are arranged on a circle.

*Results.*

Experiments have been performed to verify and evaluate our scheme. Figure.3 illustrates the experimental setup, in which, the heralded single photon source and time correlated single photon counting (TCSPC) are employed. A pulsed laser (Alnair PFL200) with central wavelength of ~1552 nm serves as degenerate pump to generate SFWM in dispersion shifted fiber (DSF) [22]. The temperature of DSF is cooled to 77 K with liquid nitrogen to reduce the Raman scattering noise. The time correlated signal and idler photons are filtered out by dense wavelength division multiplexing (DWDM) filters. The idler photons (1555.7 nm) are directly collected by InGaAs single photon detector (IDQ220), which heralds the detection of signal photons (1549.3 nm) after quantum operation. The signal photons are collimated to free space Gaussian mode expressed as $|\varphi_0\rangle$ under discrete spatial mode basis $|\varphi\rangle\langle\varphi|$. A beam splitter noted as $|\alpha\rangle\langle\varphi_0|$ is programmed on an additional spatial light modulator (Holoeye Pluto) labeled SLM0 to generate the initial state $|\alpha\rangle = \sum a_n |\varphi_n\rangle$. The target operation $|\beta\rangle = T|\alpha\rangle$ is performed by SLM1 and SLM2 (operating in the reflection mode). The projection values of final state in the spatial mode basis after the quantum operation, $|\langle\varphi_n|\beta\rangle|^2$, are detected by a single photon avalanche detector (SPAD) with mono-mode fiber coupler one-by-one. It should be mentioned that the SLM0 could be replaced by SPDC for future experiment and the spatial sampling of SPDC cone as well as beam splitting could be done simultaneous by SLM1.

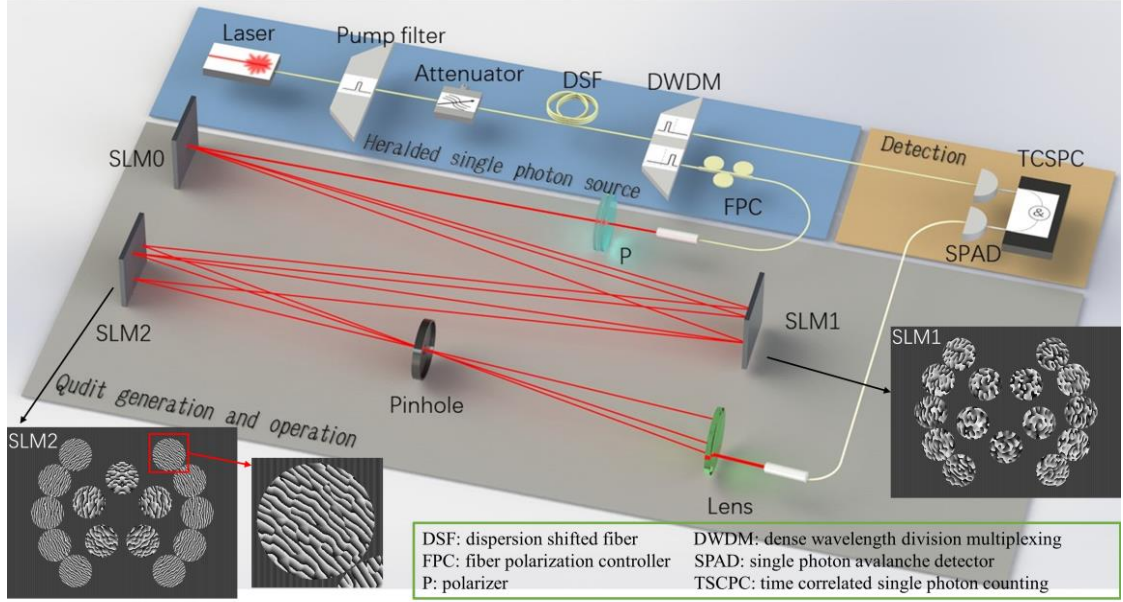

Fig.3 Experimental setup. The insets include modulation functions on SLM1 and SLM2, together with an enlarged figure of a typical phase grating.

Firstly, a $15\times15$ QFT is performed, which is an important unitary quantum operation in quantum information processing such as Shor's factorization algorithm [23]. The beam splitting gratings and beam recombining gratings encoded on SLM1 and SLM2 are shown in the insets of Fig. 3, respectively. The phase components in experimental QFT matrix $F_{\exp}$ (target QFT matrix labeled $F$) could not be measured directly in the computational basis $|\varphi\rangle\langle\varphi|$, thus the fidelity value of QFT is evaluated in the conjugate Fourier basis $|\omega\rangle\langle\omega|$:

$$|\omega_n\rangle = \sum_{d=1}^{N}\exp[-2\pi i(d-1)n/N]|\varphi_d\rangle/\sqrt{N}. \qquad (4)$$

If the initial state $|\omega_n\rangle$ is prepared precisely, $F|\omega_n\rangle = |\varphi_n\rangle$. Due to the orthogonal nature of conjunctive Fourier basis ($|\langle\omega_i|\omega_j\rangle|^2 = \delta_{ij}$), the fidelity value between experimental and theoretical QFT matrices can be expressed as $Fide(F_{\exp}, F) = Fide(F_{\exp}\Omega, F\Omega)$, where $\Omega$ is a matrix whose n-th column is $|\omega_n\rangle$. In theory, $F\Omega$ is the identity matrix acting in the computational basis $|\varphi\rangle\langle\varphi|$ and so should be $F_{\exp}\Omega$ if $F_{\exp}$ is sufficiently close to $F$. The experimental results for $F_{\exp}\Omega$ are displayed in Fig. 4 (a), which consists of coincidence counts in 120 seconds. The error bars are one standard deviation estimated from Poissonian counting statistics. The fidelity is calculated as $0.85\pm0.02$. The deviation from unit fidelity is caused mainly by dark count rates of the SPADs and the calibration error of the experimental setup. A detailed error analysis is contained in [24].

The spatial entanglement resources from SPDC can be further explored with our scheme, in which the transverse coordinates of spatial encoding basis are considered as distributed on one single circle. An equivalent demonstration of order-finding routine in Shor's factorization algorithm is shown in Fig.4 (c), where high-dimensional QFT and compiled modular exponential [] are applied on twin beams from SPDC, respectively. Furthermore, the spatial entangled state from type I SPDC (inset of Fig. 4 (c)) is one of the high-dimensional generalized Bell-states and with only one unitary matrix acting on one photon, the Bell-state can be switched to any another, for applications such as quantum teleportation of qudits. The details about such two experimental proposals based on our linear operation scheme are shown in the Supplementary Material.

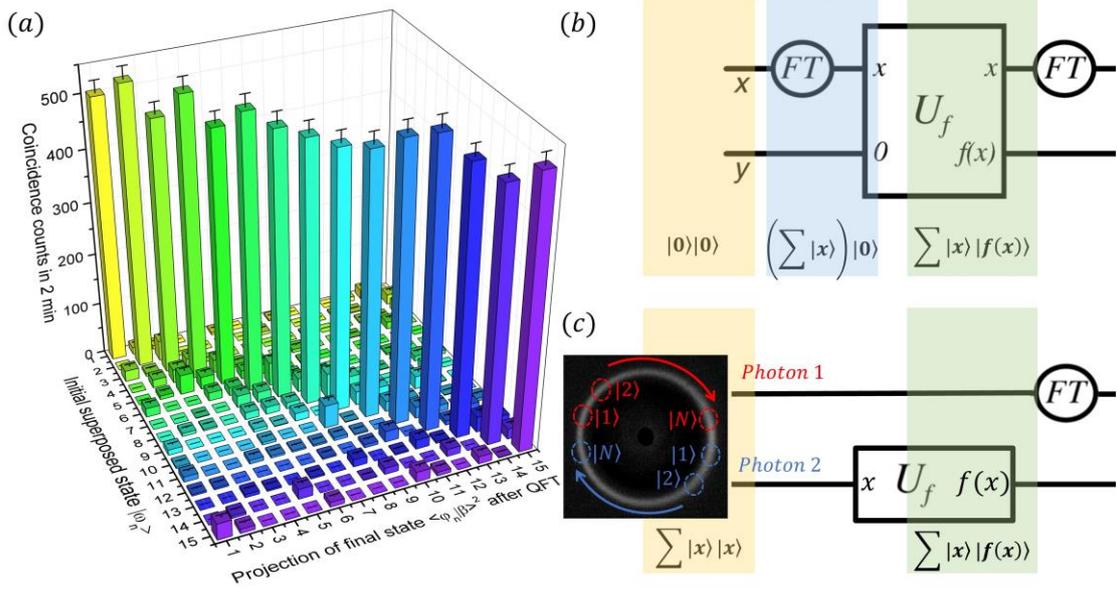

Fig.4 (a) QFT with dimensionality of $15\times15$ tested with the conjugate Fourier basis. (b) The order-finding routine of Shor's factorization algorithm. f(x) denotes modular exponential. (c) Equivalent order-finding routine with high-dimensional entangled photons. Inset, type I SPDC cone photographed by near-infrared CMOS camera.

Quantum state tomography (QST) provides a full description of the quantum state [25]. A 15-dimensional QST is performed to explore the universality of our proposal. In experiments, symmetric informationally complete positive operator-valued measures (SIC POVMs) [26] are employed. A SIC POVM basis can be generated by applying $N^2$ displacement operators on a fiducial vector [27]. After randomly choosing 100 out of totally 225 SIC POVMs, QST is performed via compressed sensing method [28]. Fig.5 are experimental results according to the representative eigenstate $|\varphi_4\rangle$ and superposed state $|\omega_2\rangle$ with definition of Eq. (4). The expectation values of projective measurements for such two states are displayed in Fig.5 (a) and (b), where filled and empty histograms indicate the recorded coincidence counts in 60 seconds and theoretical calculated references, respectively. The implemented accuracy are evaluated by statistical fidelity values, which is defined as $F_s(p_{exp}, p) = \sum \sqrt{p_{exp} \cdot p}$ [29], between theoretical and experimental probability distributions corresponding to the histograms in Fig.5 (a) and (b). The values of these two datasets are found to be 0.98 and 0.96. The corresponding density metrics (DMs) reconstructed from the randomly chosen SIC POVMs, are plotted in Fig.5 (c) and (d), in which the theoretical DMs are also displayed as empty bars. The fidelity of density matrix (DM) is evaluated with the formula [30]:

$$Fide(\rho_{exp}, \rho) = \left| Tr \sqrt{\sqrt{\rho} \rho_{exp} \sqrt{\rho}} \right|^2, \tag{5}$$

where $\rho_{exp}$ and $\rho$ are reconstructed and reference DMs, respectively. The fidelity values are 0.853 and 0.815 for eigenstate $|\varphi_4\rangle$ and superposition state $|\omega_2\rangle$, respectively. Actually, it is convenient to implement complex operators under our scheme, allowing the reconstruction of a complicated DM with 225 nonzero complex elements. Further to the results shown in Fig.5, another three similar experiments were performed. The averaged fidelity value is $0.97 \pm 0.02$ among totally 500 experimental generated 15-dimensional projective operators randomly chosen from SIC POVMs. These projective operators exhibit various amplitude and phase distributions. These results suggest that our method is valid in arbitrary linear quantum operations up to 15 dimensions and it is the first implementation of SIC POVMs on 15-dimension photonic qudits to the best of our knowledge [26].

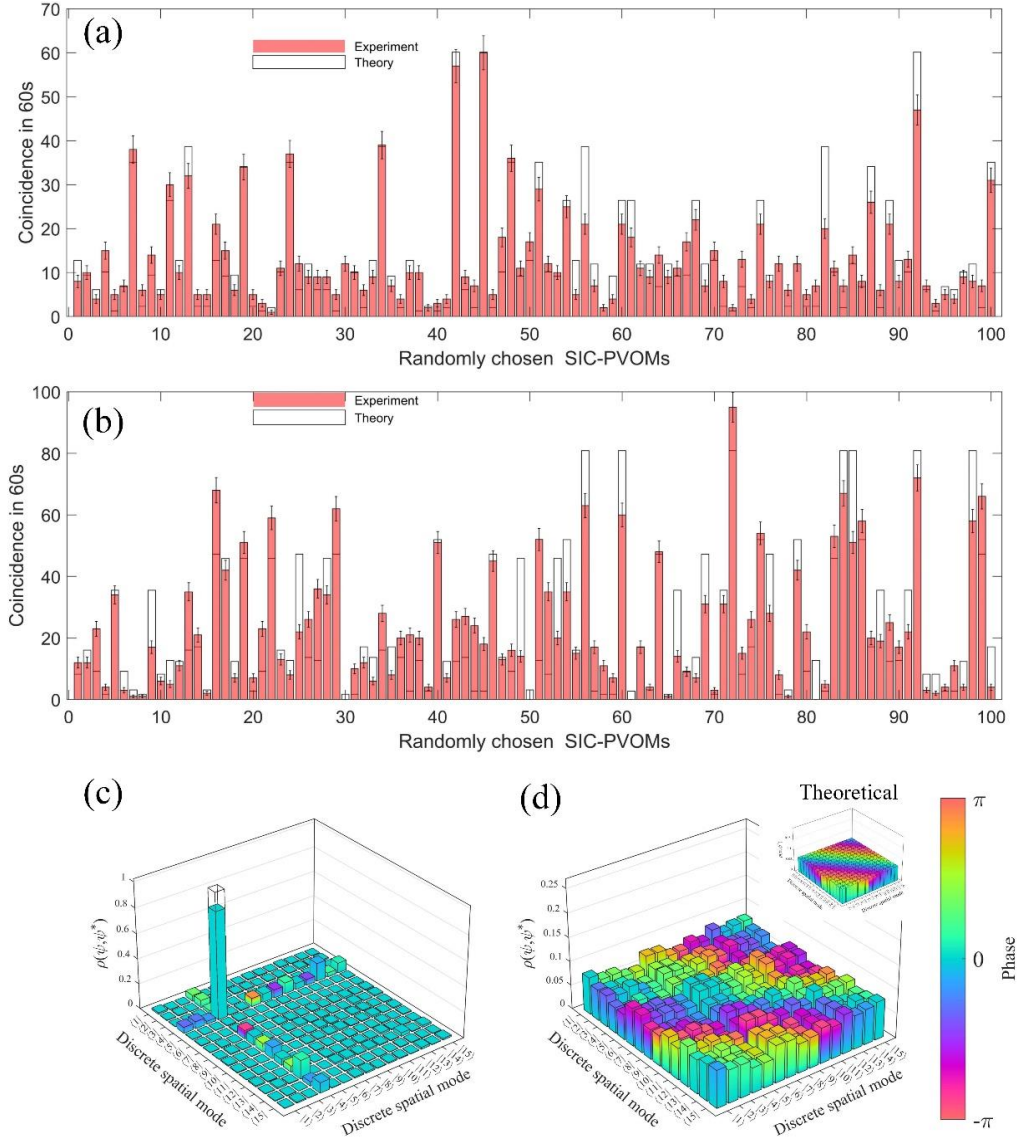

Fig.5 Quantum state tomography via compressed sensing method. (a) and (b) Projective measurements. (c) and (d) Reconstructed density matrices.

The required projective measurements can be greatly reduced with the compressed sensing technique. However, the reconstruction uncertainties would increase with high sampling ratio [28]. Figure.6 illustrates the measured fidelity and trace distance of the reconstructed DMs versus sampling ratio. Statistical results of 5 experimental data sets are plotted. The trace distance is calculated as $T(\rho_{\exp}, \rho) = Tr\left[\sqrt{(\rho_{\exp} - \rho)^\dagger (\rho_{\exp} - \rho)}\right]/2$. As the generated states under test are nearly pure, the trace distance is close to the upper bound of $\sqrt{1 - Fide(\rho_{\exp}, \rho)^2}$.

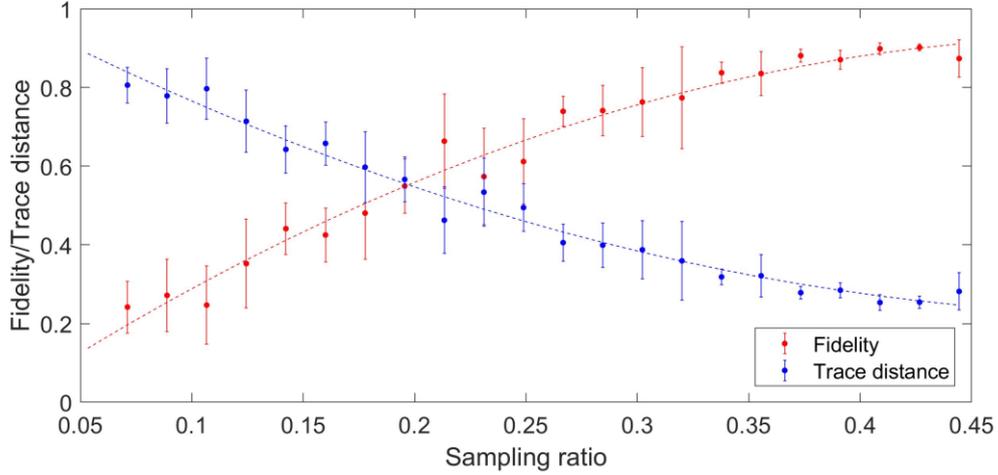

Fig.6 Fidelity and trace distance of reconstructed density matrix versus sampling ratio.

*Discussion and conclusions.*

Before summarizing, we would like to provide some specific discussions in terms of efficiency, reliability and dimensionality about our scheme. An energy factor of $1/N$ would be induced by spatial filtering with pinhole for the worst case of a $N \times N$ demonstration as previously discussed [19]. The intrinsic loss is determined by the average number of nonzero elements in each row of target matrix. Our scheme would achieve unitary efficiency for shift and clock matrices that could be used to construct any unitary matrices [6]. In the $15 \times 15$ QFT demonstration, the total insertion loss is measured as 21.2 dB. It includes the theoretical loss of ~13.7 dB, estimated by the $1/N$ factor multiplying beam splitting efficiency, and an additional loss of 7.5 dB due to the modulation efficiency and reflection rate of two SLMs. The intrinsic loss of $1/N$ is the penalty to avoid the cascaded structure. Against this, for our simple non-cascading structure, the attenuation induced by the optical elements should be largely independent of the dimensionality of linear operator realized. Moreover, our scheme exhibits robustness against phase modulation errors compared with a cascaded design [19].

The quantum operators realized here were demonstrated with dimensionality of 15, but an extension of the 24-dimensional arbitrary linear transformation has been demonstrated previously [19]. In principle the achievable dimensionality is only limited by the achievable spatial resolution of phase modulation. Actually, our scheme is not limited by SLMs and any passive holographic elements to perform the phase modulation would be feasible, *e.g.* metasurfaces [31]. Thus, it is potential to achieve on-chip high-dimensional quantum operator with our scheme.

In summary, we have proposed and demonstrated a simple and flexible scheme for universal quantum operations with high fidelity and high dimensionality. Two linear quantum operations of QFT and SIC POVMs have been performed with dimensionality of $15 \times 15$. Since the transverse spatial modes are employed, we believe that our work would be potential to fully explore SPDC on high-dimensional quantum applications including Bell-states switching and compiled demonstration of Shor's factorization algorithm.

Acknowledgement:

This work was supported by the National Key Research and Development Program of China (2017YFA0303700, 2018YFB2200402), the National Natural Science Foundation of China (Grant No. 61875101 and 61621064), Beijing Innovation Center for Future Chip and Beijing academy of quantum information science. SL would like to thank Peng Zhao, Xin Yao and Rong Xue for their valuable discussions and helpful comments. In addition, SMB thanks the Royal Society for support through the grants RP/150122 and RP/EA/180010.

Supplementary material for:
Programmable coherent linear quantum operations with high-dimensional optical spatial modes

*Shikang Li[1], Shan Zhang[1] Xue Feng[1,]\*, Stephen M. Barnett[2], Wei Zhang[1], Kaiyu Cui[1], Fang Liu[1], Yidong Huang[1]*

\*Corresponding Author: E-mail: x-feng@tsinghua.edu.cn

[1]Department of Electronic Engineering, Tsinghua University, Beijing 100084, China.
[2]School of Physics and Astronomy, University of Glasgow, Glasgow G12 8QQ, United Kingdom.

## I. SPATIAL LIGHT MODULATORS ENCODING METHOD

The spatial resolution of the employed spatial light modulator (SLM, Holoeye Pluto series) is $1920 \times 1080$, while the size of each single pixel is about $8\mu m \times 8\mu m$. Firstly, we would like to introduce some ancilla phase modulation patterns to simplify the following explanation.

### Phase modulation functions

As mentioned in the main text, the modulation efficiency of SLMs is less than unity. To separate effectively the components directly reflected without modulation, first order diffraction is employed, and the function of phase modulation reads

$$F_{grating}(\boldsymbol{r}) = \exp(i\boldsymbol{k_{gratin}} \cdot \boldsymbol{r}) \tag{1}$$

With the help of such blazed gratings, the directly reflected components can be spatially filtered out. In this work, the transverse wave vector $\boldsymbol{k_{grating}}$ is selected to obtain blazed grating with a spatial period of 4 pixels.

To maintain the transverse size of Gaussian spots against natural divergence during propagation, a symmetric confocal cavity is implemented between SLM1 and SLM2 by Fresnel lens programmed on SLMs. Under the paraxial approximation, the transmission function of a lens with a focal length of $f$ is

$$F_{lens}(\boldsymbol{r}) = \exp\left(\frac{ik|\boldsymbol{r}|^2}{2f}\right) \tag{2}$$

Additionally, a binary function is employed to avoid spatial coincidences of phase gratings acting on different Gaussian spots $|\varphi_n\rangle$. The expression is

$$\chi(\boldsymbol{r}) = \begin{cases} 1, & |\boldsymbol{r}| < R_{threshold} \\ 0, & elsewhere \end{cases} \tag{3}$$

The value of $R_{threshold}$ is determined according to the beam waist of a single Gaussian spot. The amplitude modulation is implemented by a checker-board method. The phase modulation settled on SLM0 is

$$F_{diff0}(\boldsymbol{r}) = exp\left\{iarg\left[\sum_{n=1}^{N} \xi_n a_n \exp(i\boldsymbol{k_n} \cdot \boldsymbol{r})\right]\right\} F_{lens}(\boldsymbol{r})\chi(\boldsymbol{r})F_{grating}(\boldsymbol{r}) \tag{4}$$

where $\{a_n\}$ is the state vector of the initial state, and $\{\xi_n\}$ are optimization parameters to ensure desired beam splitting. The transverse wave vectors of $\{k_n\}$ determine the propagation directions of initial Gaussian spots and they are utilized to mimic the spontaneous parametric down conversion cones produced in a nonlinear crystal. Under the paraxial approximation, the definition of $k_n$ is

$$k_n = \frac{kR_n}{2f}, \forall n = 1\ldots N \tag{5}$$

The transverse coordinates of the *n*-th Gaussian spot is $R_n$ as mentioned in the main text. The phase modulation on SLM1 is chosen to be

$$F_{diff1}(r) = \sum_{m=1}^{N} exp\left\{iarg\left[\sum_{n=1}^{N} \mu_{mn} A_{mn} \exp(i(k_{mn} - k_m) \cdot [r - R_m] - i\theta_{mn})\right] - i\delta_m\right\} \cdot F_{lens}(r - R_m)\chi(r - R_m) \cdot F_{grating}(r) \tag{6}$$

The parameters of $\mu_{mn}, A_{mn}$ and $k_{mn}$ in Eq. (6) have the same definition as those in the main text. As the optical path of each spot is different during propagating from SLM1 to SLM2, the phase compensation of $\theta_{mn}$ should also be different for each splitting direction and can be expressed as

$$\theta_{mn} = \frac{k|R_n - R_m|^2}{4f} \tag{7}$$

Similarly, different phase compensation of $\delta_m$ is employed for each initial Gaussian spot to compensate the path difference during propagating from SLM0 to SLM1. The value of $\delta_m$ is settled as

$$\delta_m = \frac{k|R_m|^2}{4f} \tag{8}$$

In addition to beam splitting and focusing lens, extra beam refraction functions are also employed in Eq. (6) to compensate for the initial transverse wave vectors $\{k_n\}$ of individual Gaussian spots generated by SLM0.

The function of SLM2 is beam recombining. As is it the reverse procedure of beam splitting, the phase modulation functions are similar to those on SLM1. The phase modulation settled on SLM2 is

$$F_{diff2}(r) = \sum_{m=1}^{N} exp\left\{iarg\left[\sum_{n=1}^{N} \nu_{mn} B_{mn} \exp(-ik_{mn} \cdot [r - R_m])\right]\right\} \cdot F_{lens}(r - R_m)\chi(r - R_m) \cdot F_{lens}(r)F_{grating}(r) \tag{9}$$

The parameters of $\nu_{mn}, B_{mn}$ and $k_{mn}$ in Eq. (9) have the same definition as those in the main text. Different from the phase modulation on SLM1, there is no need for extra additional refraction and phase compensation on SLM2, but an extra lens is required for the last step, which is beam filtering with a pinhole. A pinhole with transmission function described by $\chi(r)$ is positioned one focal length of $f$ after SLM2. To accomplish beam filtering, a lens with focal length of $f$ is placed at one focal length away after pinhole, as presented in the experimental setup in main text.

Phase calibration method

In the laboratory implementation, phase errors arise due to the misalignment of optical elements. In addition, there would be extra phase errors if the phase compensation terms in Eq. (6) are not estimated precisely. Fortunately, these phase errors are constant and independent of the target matrices. Thus, the phase errors can be corrected before experiments. The amplitudes of matrix elements are robust against misalignments and there is no need for amplitude calibration in most situations.

The phase calibration is performed with a laser source (RIO ORION) centered at 1550 nm in place of the heralded single photon source in experimental setup. The final state after linear operation is measured by a charge coupled device (CCD) camera. Generally, the influences of constant phase error $\varepsilon$ on the implemented matrix transformation $T^{real}$ turn out to be

$$T_{mn}^{real} = T_{mn} \exp(i\varepsilon_{mn}) \tag{10}$$

The error matrix $\exp(i\varepsilon_{mn})$ is measured measured so that the phase errors can be compensated by implementing $T_{mn} \exp(-i\varepsilon_{mn})$ rather than $T_{mn}$. Any matrix with no zero elements could be adopted for phase calibration. In our experiments, the discrete Fourier transformation (DFT) is performed without phase calibration at first. Then, the phase terms of all elements in the realized DFT are measured. Finally, the error matrix $\exp(i\varepsilon_{mn})$ is deduced by comparing the achieved DFT and the standard DFT. A tomographic method is employed to measure the phase terms. By measuring the amplitudes of output vectors corresponding to meticulously designed input vectors, the phase terms of transformation matrix could be deduced. Here, the principle of calibration is shown with an example of $3 \times 3$ matrix.

$$\begin{bmatrix} O_{11} & O_{12} & O_{13} & O_{14} \\ O_{21} & O_{22} & O_{23} & O_{24} \\ O_{31} & O_{32} & O_{33} & O_{34} \end{bmatrix} = \begin{bmatrix} T_{11}^{real} & T_{12}^{real} & T_{13}^{real} \\ T_{21}^{real} & T_{22}^{real} & T_{23}^{real} \\ T_{31}^{real} & T_{32}^{real} & T_{33}^{real} \end{bmatrix} \begin{bmatrix} 1 & 1 & i & i \\ 1 & 0 & 1 & 0 \\ 0 & 1 & 0 & 1 \end{bmatrix} \tag{11}$$

The interference results of matrix elements in first column with elements in other columns are available as the intensities in output vectors. To calculate the relative phase with respect to elements in first column precisely, the inner product is performed twice. For example, the phase of $T_{12}^{real}$ can be found from

$$I_{12}^{cos} = |O_{11}|^2 = \left|T_{11}^{real} + T_{12}^{real}\right|^2, \quad I_{12}^{sin} = |O_{13}|^2 = \left|iT_{11}^{real} + T_{12}^{real}\right|^2 \tag{12}$$

As the amplitudes of elements in $T^{real}$ are known, the relative phases can be deduced from Eq. (12) without any ambiguity. Generally, $2(N-1)$ tests with different input vectors are required to measure all the relative phase terms of the $N \times N$ matrix.

There would be, however, a constant relative phase between different rows of matrix that is not calibrated. Actually, these relative phase terms can be considered as path differences from each output port of the transformation to its corresponding detector. Thus, these phase terms have no influence in quantum computing tasks, including multiphoton interference such as boson sampling.

It is possible to handle these phase differences among matrix rows, as a different method for phase measurement is introduced in our previous work [1], where a reference field is required. However, the tomographic method utilized in this work is simpler and more accurate. The test results of the calibrated DFT matrix acting on the Fourier basis are shown in Fig.1. The input and output state vectors are captured by CCD camera. All the input Fourier bases of $|\omega_n\rangle$ have uniform amplitude but different phase distributions. For different input Fourier basis, the output of DFT matrix would be single Gaussian spots in different positions. Each time only one output port is bright while other output ports vanish due to multiport interference, as they should.

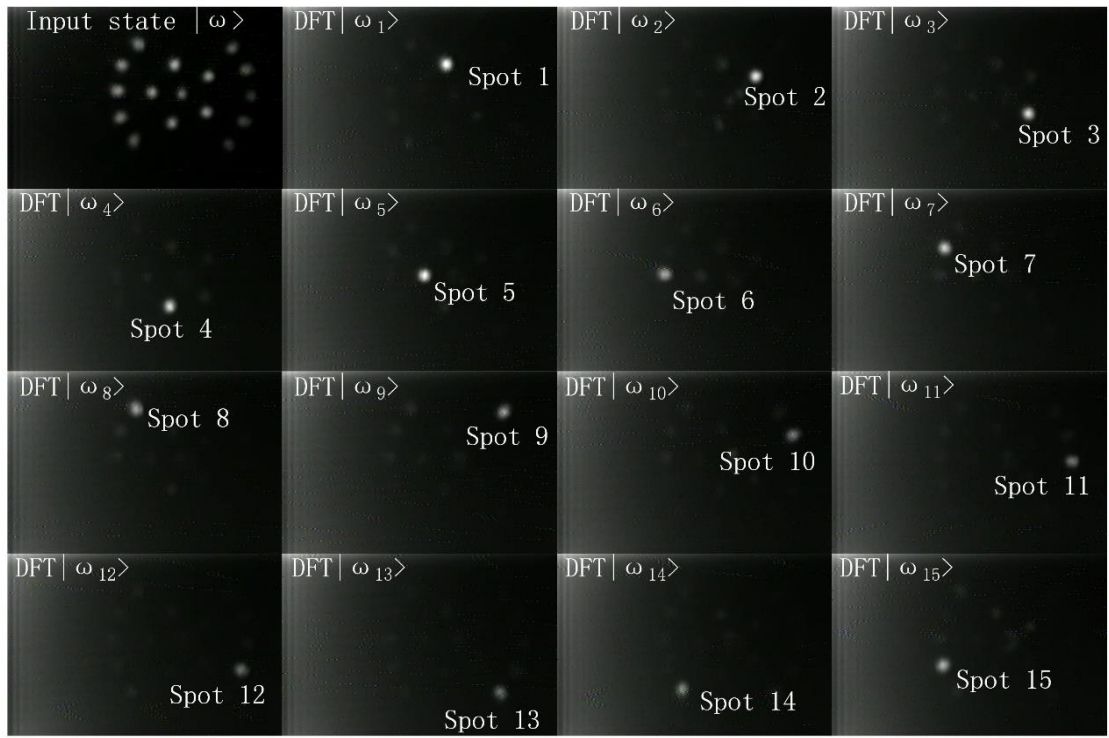

Fig.1 The results of calibrated DFT matrix under Fourier basis. For different input Fourier states, the output of DFT matrix would be single Gaussian spots in different positions.

The stability of our calibration method is tested over a period of one week. The fidelity of $15 \times 15$ DFT demonstration is selected as an example. The fidelity drops from ~ 93% to ~87% after 7 days without any more calibration. Once the calibration is performed, since the constant phase error is independent to transformation matrices, any target matrices can be readily implemented with high fidelity.

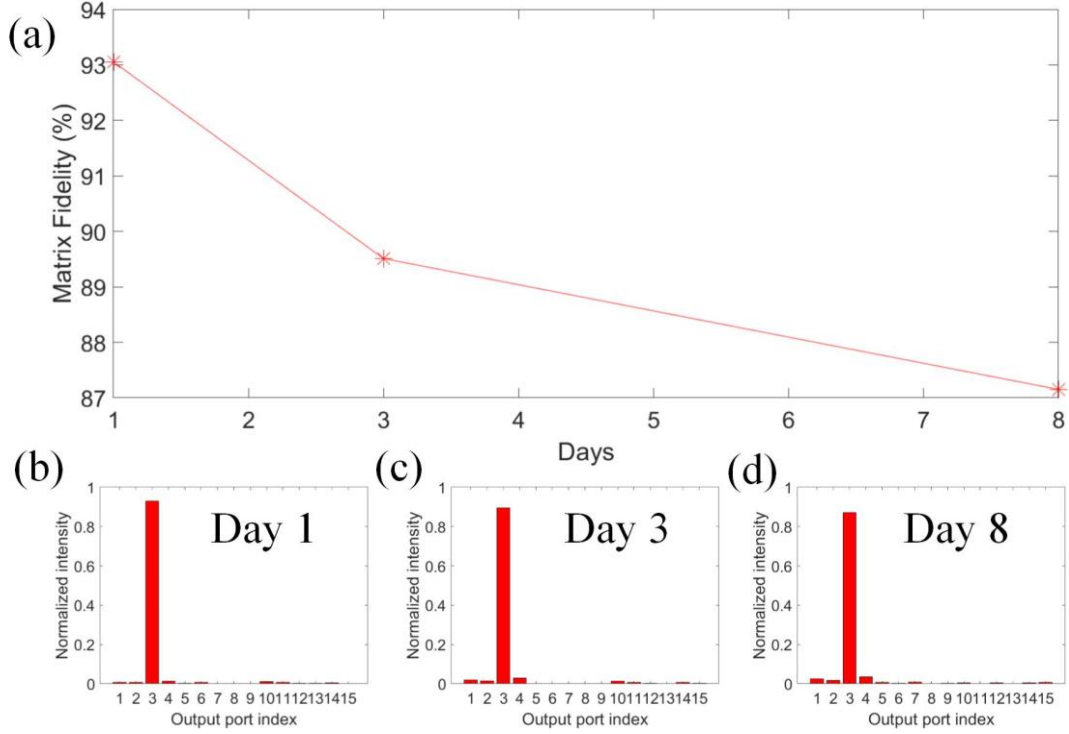

Fig.2 (a) Matrix fidelity variation of calibrated $15 \times 15$ DFT demonstration over one week. (b~d) The output vector of DFT with input Fourier base of $|\omega_3\rangle$.

## II. ERROR ANALYSIS

The errors can be categorized as following: (a) errors introduced by the time correlated single photon counting (TCSPC), (b) errors in design and implementation of modulation functions on SLMs and (c) imperfections in matrix calibration. Among them, the dominating one is introduced by the TCSPC due to the Poissonian counting statistics and dark noises in detector. The errors of the parameters in SLM modulation functions as well as imperfections in matrix calibration only lead to a little deterioration of the fidelity for the implemented linear operators and projectors.

### Errors induced by TCSPC

The raw photon counting rate of the heralded single photon source is about 270 Hz with the IDQ220 detectors. However, the total insertion loss of experimental setup is ~ 32dB including polarization control, high-dimensional state generation, linear operation and free space to fiber coupling. In addition to the intrinsic loss of 11.76 dB to implement $15 \times 15$ matrix, there is an extra loss of 9.52 dB caused by the non-unit modulation efficiency of SLM1 and SLM2. Further, extra losses of 3.01 dB, 4.46 dB and 4.15dB are induced by polarization control as well as fiber to free space collimation, state generation with SLM0, free space to fiber collection, respectively. As is shown in the main text, the data accumulating time for each measurement is one minute, hence only tens of coincidence events can be recorded. As the single photon events follow the Poissonian distribution, the uncertainty is $\sim\sqrt{N_{count}}$ for data series. When $N_{count}$ is low, the relative uncertainty of coincidence counting arises and has been proved as the main reason for fidelity deterioration of quantum projective measurements. To further confirm this, we have performed high-dimensional state tomography with the same projective method but with classical coherent light source instead. The results for symmetric informationally complete positive operator-valued measurements (SIC POVMs) as well as reconstruction of density matrices are presented in Fig.4 in section IV. For four groups of experiments, the statistical fidelity values of SIC POVMs are 0.979~0.996, while the fidelity values of density matrices are 0.930~0.994. With intensive light source, the errors decrease evidently. This suggest that the uncertainty of coincidence counting is the critical error in this demonstration.

The compressed sensing technique would further amplify the uncertainties. According to compressed sensing theory [2], the error $\varepsilon_{DM}$ of reconstructed density matrix follows

$$\varepsilon_{DM} \geq \left(\sqrt{N^2/m}\right)\varepsilon_P \qquad (13)$$

$N$ is the dimensionality of density matrix, while $m$ is the number of projective measurements. $\varepsilon_P$ represents errors of projection values. As the sampling ratio is about 0.35 in this work, the errors of density matrices would be at least ~1.7 times of errors of raw projection data, which could explain the fidelity deterioration from quantum projection values to density matrices.

The accidental coincidence rate caused by dark counts in the single photon detectors is measured to be ~2 per minute. The fidelity deterioration of DFT results in the main text is mainly caused by dark counts occurring in those output ports that should be quenched due to destructive interference of single photon.

### Errors induced by modulation functions on SLMs

In principle, the phase-only modulation functions should be modified by optimization parameters of $\mu_{mn}$ and $\nu_{mn}$ in Eq. (6) and Eq. (9) to achieve unit fidelity. However, it is highly time-consuming to optimize the phase gratings every time for different tasks and after phase calibration. Moreover, as mentioned in the main text, near unitary fidelity value can be achieved even by directly setting $\{\mu_{mn}\}$ and $\{\nu_{mn}\}$ to 1. This allows fast design of phase modulation functions to realize a general target matrix, only if the fidelity deterioration induced is negligible.

To quantify the influence of optimization parameters, we have mathematically modeled our experimental setup so that it can be simulated according to the Huygens-Fresnel theorem under the paraxial approximation. Thus, we can characterize the theoretical fidelity of matrix transformations under all the experimental conditions including beam splitting, beam recombining, free space propagation and spatial filtering. 1000 unitary matrices have been randomly generated with the dimensionality of $15 \times 15$. The phase modulation functions for these matrices are designed according to Eq. (6) and Eq. (9) with all the optimization parameters equal to 1. Fig. 3 presents the fidelity histogram of 1000 numerical simulation trials. The standard derivation of fidelity is calculated as $\delta F < 1.5 \times 10^{-4}$. The high fidelity values and small fidelity derivation suggest that the fidelity deterioration without optimization is negligible.

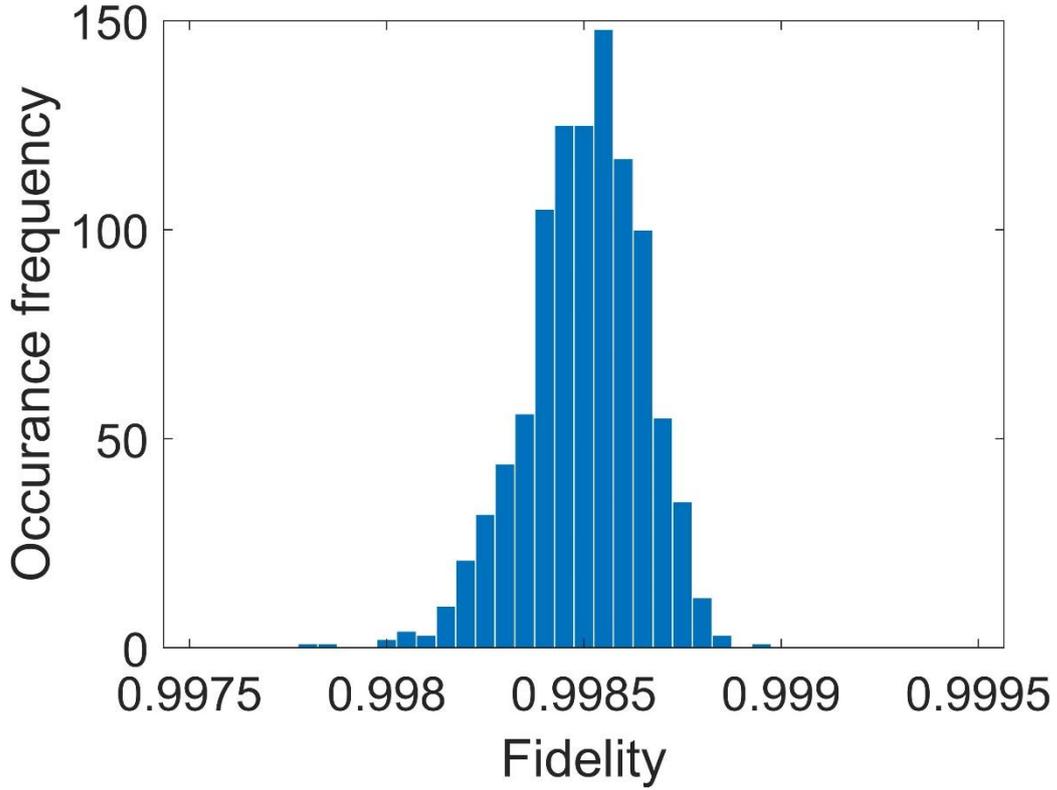

Fig.3 The histogram for normalized fidelity of different unitary matrices with dimensionality of $15 \times 15$. The phase gratings are not optimized in these designs. The histogram is obtained through Monte Carlo simulation with 1000 trials.

The optical path difference induced during propagating from SLM1 to SLM2 can be compensated in Eq. (6). However, the phase compensation is valid only when the coherence length of single photons is larger than the optical path difference. In this demonstration, the distance between SLM1 and SLM2 is 0.8 m, while the maximum and average distances between Gaussian spots in transverse plane are ~5 mm and ~1 mm, respectively. This leads to a maximum optical path difference of ~20 times of optical wavelength and an average optical path difference of about one wavelength. The above estimate brings additional requirement for coherence length of photon sources. The influences of optical path differences will be sufficiently small when the linewidths of the pump filter and signal (idler) filter are narrow enough.

Errors induced by matrix calibration

Matrix calibration is performed with laser source centered at 1550 nm, while the center wavelength of signal filter for single photons is 1555.7 nm. As SLMs are not narrow-band devices, these two wavelengths can be considered as the same for phase calibration.

During phase calibration, the amplitudes of matrix elements are supposed to be accurate. The simulated results shown in Fig. 3 supports this assumption. The intensities of output states are measured by an InGaAs CCD camera. The nonlinear effects as well as background noises of CCD camera would also induce some errors, which are mostly responsible for the fidelity difference of experimental results with intensive light presented in Fig. 4.

## III. FEASIBLE EXPERIMENTS WITH SPDC

When the transverse coordinates of spatial encoding basis are considered as distributed on one single circle, the two-photon spatial-entangled state generated via frequency-degenerate type-I SPDC can be expressed as $\sum_{x=0}^{N-1}|x\rangle|x\rangle$ with momentum conservation, where $N$ equals to the half of the sampling number on the SPDC cone. Based on this $N$-dimensional entangled initial state, there are two experimental proposal with our linear operation scheme as followed.

### Compiled demonstration of Shor's factorization algorithm

To factor a composite number $M = pq$ is equivalent to find the period $r$ of the modular exponential function (MEF) of $f(x) = a^x(mod\ M)$, where number $a$ can be chosen to any integer. Shor's algorithm provides an effective routine [3-5] to find the period of $r$, where the "quantum parallelism" can be achieved by coherent manipulations and detections of the highly entangled states:

$$\frac{1}{\sqrt{N}}\sum_{x=0}^{N-1}|x\rangle|a^x(mod\ M)\rangle \tag{14}$$

The value of $N$ denotes the size of each quantum register set. Next, the quantum Fourier transformation (QFT) of dimensionality $N$ is applied on the first register, yielding to quantum interference from which information about period $r$ of $f(x)$ can be deduced [5]:

$$\frac{1}{N}\sum_{y=0}^{N-1}\sum_{x=0}^{N-1}e^{2\pi ixy/N}|y\rangle|a^x(mod\ M)\rangle \tag{15}$$

Thus, the preparation of states in eq. (15) is essential for demonstrating Shor's algorithm.

The initial states of two quantum registers are separable ground states in the original order-finding routine [3]. Eq. (14) is achieved by multiphoton execution that entangles $N$ input value of $x$ in first register with corresponding value $f(x)$ in the second register. However, with initial entangled state $\sum_{x=0}^{N-1}|x\rangle|x\rangle$ in our proposal, there is no need for generating entanglement with multiphoton controlled gates, which would lead to low energy efficiency with post selection [6]. The QFT is applied on the first quantum register, yielding

$$\frac{1}{N}\sum_{y=0}^{N-1}\sum_{x=0}^{N-1}e^{2\pi ixy/N}|y\rangle|x\rangle \tag{16}$$

Next, the MEF is applied on the second quantum register. Thus, the state in eq. (15) is prepared and the quantum parallelism could be exhibited by simultaneously readout of two registers. Though MEF is nonlinear operation and could not be implemented directly with any linear operation scheme, a proof-in-principle experiment of Shor's algorithm would be feasible with linear transformation performing the pre-compiled MEF [5]. As a concrete example, with $N = 16$ for number of sampling on SPDC cone and the dimensionality of QFT, the factorization of $15 = 3 \times 5$ could be demonstrated. Additionally, such implementation is not limited to the "easy" case [5,7] of $a = 11$ that has been previously reported with photonic platform.

It is significant that high-dimensional entangled state is employed as the initial state, thus multiphoton controlled gates are avoided. Due to the high-dimensional encoding, only one photon is needed for each quantum register, thus the coherent manipulation and detection could be further simplified.

## Generation of complete high-dimensional Bell basis

The complete high-dimensional Bell basis can be generated with our proposal. The $N$-dimensional Bell basis can be written as [8,9]:

$$|\psi\rangle_{mn} = \frac{1}{\sqrt{N}} \sum_{x=0}^{N-1} e^{2\pi i x n/N} |x\rangle |mod(x+m, N)\rangle \tag{17}$$

Begin with the initial state sampled from the SPDC cone, we have one of the $N$-dimensional Bell states:

$$|\psi\rangle_{00} = \frac{1}{\sqrt{N}} \sum_{x=0}^{N-1} |x\rangle |x\rangle \tag{18}$$

With $N$-dimensional shift matrices and clock matrices applying on the second photon, the Bell state of $|\psi\rangle_{00}$ can be switched to $|\psi\rangle_{m0}$ and $|\psi\rangle_{0n}$, respectively. Here, shift matrices and clock matrices are expressed as:

$$T_{shift,m} = \frac{1}{\sqrt{N}} \sum_{x=0}^{N-1} |mod(x+m, N)\rangle\langle x| \tag{19}$$

$$T_{clock,n} = \frac{1}{\sqrt{N}} \sum_{x=0}^{N-1} e^{2\pi i x n/N} |x\rangle\langle x| \tag{20}$$

The state of $|\psi\rangle_{00}$ can be switched to any target Bell state $|\psi\rangle_{mn}$ by applying the production of $T_{shift,m}$ and $T_{clock,n}$ on the second photon. Thus, the complete $N$-dimensional Bell basis can be generated by type-I SPDC combined with our linear operation scheme. Furthermore, unitary energy efficiency can be achieved when performing shift matrices and clock matrices, as well as any of their productions. As a concrete example, 15-dimensional complete Bell basis could be generated with our proposed $15 \times 15$ arbitrary matrix transformations. As the maximally entangled two-particle quantum states, the complete Bell basis could be employed for high-dimensional quantum protocols such as teleportation, dense coding, and entanglement swapping.

## IV. ADDITIONAL LABORATORY RESULTS

The experimental results of state tomography with classical laser source are presented in Fig. 4. Additional data for quantum state tomography that are not included in the main text are displayed in Fig. 5. Fig. 4 presents four groups of experimental results. The three columns in Fig. 4 indicate results of SIC POVMs, experimental density matrices, theoretical density matrices, respectively. Fig. 4(a) represents characterization of eigenstate, while Fig. 4(b~d) are those of superposed states. For Fig.4 (a~d), the statistical fidelity values of SIC POVMs are 0.996, 0.985, 0.979 and 0.981, while the fidelity values for state tomography are 0.994, 0.946, 0.930 and 0.940.

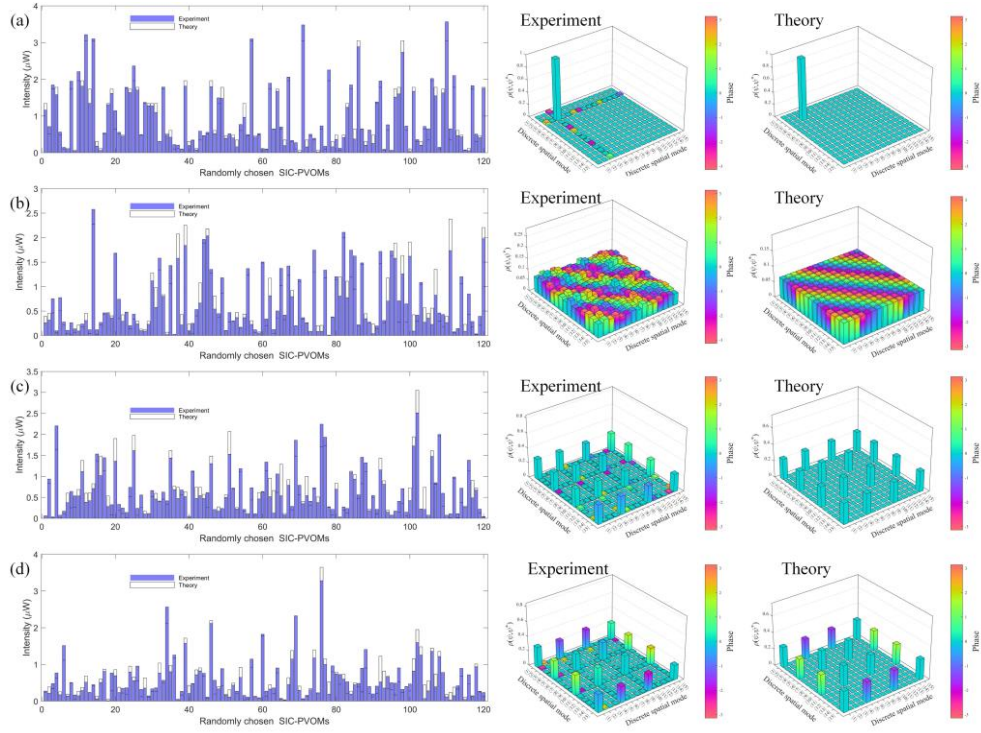

Fig. 4 Four groups of state tomography with classical coherent light source. Three columns are corresponding to the results of SIC POVMs, experimental density matrices, and theoretical density matrices, respectively. (a) Results of eigenstate. (b~d) Results of superposed states.

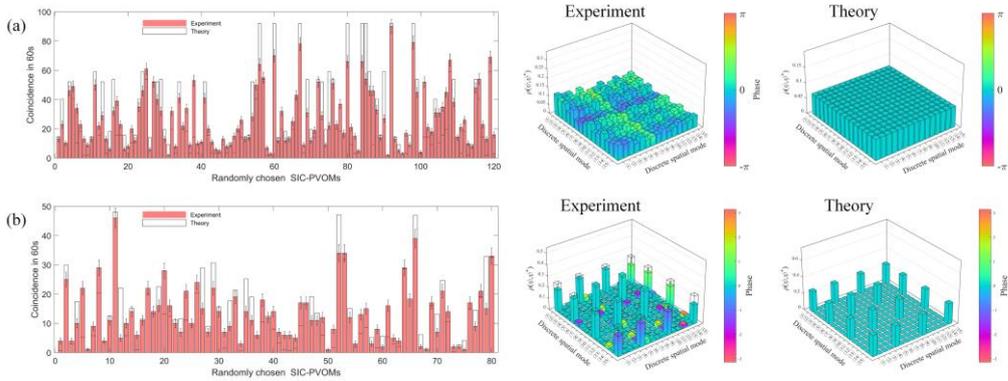

Fig. 5 Two groups of quantum projective measurements and quantum state tomography. Three columns are corresponding to the results of SIC POVMs, experimental density matrices, and theoretical density matrices, respectively. The statistical fidelity values of SIC POVM are 0.934 and 0.936 for (a) and (b), respectively. The fidelity values of quantum state tomography are 0.806 and 0.738.